\documentstyle[12pt,epsf]{article}

\newcommand \blackboardrrm{\mathchoice
{\rm I\kern-0.21 em{R}}{\rm I\kern-0.21 em{R}}
{\rm I\kern-0.19 em{R}}{\rm I\kern-0.19 em{R}}}

\newcommand \blackboardzrm{\mathchoice
{\rm Z\kern-0.32 em{Z}}{\rm Z\kern-0.32 em{Z}}
{\rm Z\kern-0.28 em{Z}}{\rm Z\kern-0.28 em{Z}}}

\newcommand \be  {\begin{equation}}
\newcommand \ee  {\end{equation}}
\newcommand \lan {\langle}
\newcommand \ran {\rangle}

\newcommand{\overl}[1]{\overline{#1}}

\begin{document}

\title{Crossovers in the Two Dimensional Ising Spin Glass with ferromagnetic
next-nearest-neighbor interactions}

\author{Giorgio Parisi$^a$, \,\, Juan J. Ruiz-Lorenzo$^a$\, 
and \, Daniel A. Stariolo$^b$\\[0.5em]
$^a$ {\small Dipartimento di Fisica and INFN, Universit\`a di Roma}
   {\small {\em La Sapienza} }\\
{\small   \ \  P. A. Moro 2, 00185 Roma (Italy)}\\[0.3em]
$^b$  {\small Departamento de Fisica, Universidade Federal de Vi\c{c}osa}\\
{\small 36570-000 Vi\c{c}osa, MG (Brazil)}\\[0.3em]
{\small \tt giorgio.parisi@roma1.infn.it }\\
{\small \tt ruiz@chimera.roma1.infn.it }\\
{\small \tt            stariolo@mail.ufv.br}\\[0.5em]}

\date{January 21, 1998}

\maketitle

\begin{abstract}

By means of extensive computer simulations we analyze in detail the
two dimensional $\pm J$ Ising spin glass with ferromagnetic
next-nearest-neighbor interactions. We found a crossover from
ferromagnetic to ``spin glass'' like order both from  numerical
simulations and  analytical arguments. We also present evidences
of a second crossover from the ``spin glass'' behavior to a
paramagnetic phase for the largest volume studied.

\end{abstract}  

\thispagestyle{empty}
\newpage

%+++++++++++++++++++++++++++++++++++++++++++++++++++++++++++++++++++++++++++
\section{\label{S_INT}Introduction}

At present it is clear that the lower critical dimension of
Edwards-Anderson Ising spin glasses is at some point between $D=2$ and
$D=3$ \cite{BOOK}. By using different numerical techniques as for
instance Monte Carlo simulations, exact ground state calculations and
diagonalization of the transfer matrix, it is rather well established
that the two dimensional Ising spin glass presents a transition at
$T=0$.  Furthermore, an experimental realization of a 2D spin glass
also points to a transition only at $T=0$~\cite{dekker,mydosh}.
However the values of the critical exponents are still a matter of
controversy\cite{BOOK,review,bhatt,kawashima,rieger,mydosh}\footnote{Recently
has been pointed out by some authors~\cite{SHIRAKURA} that the two
dimensional Ising spin glass shows a phase transition at finite
temperature, nonetheless the scaling plots of these works are
compatible with a $T=0$ phase transition in agreement with all the
previous cited studies.}.

In a recent paper Lemke and Campbell \cite{lemke} studied the $2D \pm
J$ Ising spin glass with nearest neighbors interactions adding a
ferromagnetic interaction between
next-nearest-neighbors. Surprisingly, they found numerical evidence
for a spin glass transition at finite temperature simulating
intermediate lattice sizes.

In order to gain more insight on the nature of the low temperature
properties of the model we have computed through extensive simulations
a variety of global quantities as well as staggered ones
(i.e. observables defined in one of the two sub-lattices in which the
system naturally divides), like spin glass and ferromagnetic
susceptibilities, specific heat, Binder cumulants, magnetizations and
the spin glass order parameter.

In particular we have studied how the originally stable staggered
phase is destroyed by the spin glass interaction and finally we have
studied if the spin glass phase is stable in the thermodynamic limit.

We will show numerical evidences for two different crossovers: the
first one from the staggered ferromagnetic phase to a spin glass
phase; and the second one from the spin glass phase to a paramagnetic
phase.

The plan of this paper is the following. In the next section we will
define the model and describe limiting cases in the parameter
space. In the third section we will map the model to the Random Field
Ising Model and we will obtain analytical evidences of both
crossovers (moreover we will compute the dependence of the first
crossover length on the parameters of the model). In sections fourth
and fifth we will study numerically the model. Finally we will present
our conclusions.

%+++++++++++++++++++++++++++++++++++++++++++++++++++++++++++++++++++++++++++
\section{\label{S_MOD}The model}

The model we are considering is defined by the following Hamiltonian
on a square lattice with periodic boundary conditions
\be
H= -\sum_{<ij>} J_{ij} S_i S_j - K \sum_{\ll k l \gg} S_k S_l \ ,
\label{model}
\ee 
where $<ij>$ denotes sum over all the first nearest neighbors pairs
(distance 1) and $\ll k l\gg$ denotes sum over all the second nearest
neighbors pairs (distance $\sqrt 2$).The couplings $J_{ij}$ are quenched
variables with $J_{ij}=\pm \lambda$ with probability $1/2$. $\lambda$
and $K$ are positive constants whose ratio determines the relative
strength between spin glass and ferromagnetic interactions. In the
rest of the paper we fix $K=1$. In order to fix the notation we also
define the variance of the $J_{ij}$ distribution as $\sigma^2_J \equiv
\overline{ J_{ij}^2}=\lambda^2$.

We next describe two limiting cases of the model (for $K=1$):
\begin{itemize}

\item For $\lambda \to \infty$ the system is the $2D$ spin glass for
which is known that $T_c=0$
\cite{BOOK,review,bhatt,kawashima,rieger,mydosh}.

\item When $\lambda=0$ the whole lattice decouples into two independent
sub-lattices (that we will denote hereafter as sub-lattices 1 and 2,
the black and white sub-lattices in a chess-board. 
Each sub-lattice is itself a two
dimensional ferromagnetic Ising model and will have a phase transition
(paramagnetic-ferromagnetic) just at the Onsager temperature: $T_c
=2/\log(1+\sqrt 2) \simeq 2.269$. The order parameter of the
phase transition is the so-called staggered magnetization: i.e. the
magnetization of one of the two sub-lattices. Obviously, the
probability distribution of the total magnetization of the whole
lattice for low temperatures will take into account the four possible
different magnetizations of the two independent sub-lattices.

\end{itemize}

So initially we have, when $\lambda=0$, ferromagnetic order (in
both sub-lattices), that  can be affected by the introduction of
spin glass couplings. The effect of these couplings is to couple (by
means of a spin glass interaction) both sub-lattices.

In the next section we will examine analytically the effect of the
introduction of random couplings (linking both sub-lattices) 
in the original stable (staggered) ferromagnetic order.

%+++++++++++++++++++++++++++++++++++++++++++++++++++++++++++++++++++++++++++
\section{\label{A_RES}Analytic Results}

We will study in this section the analogy, suggested by 
Lemke and Campbell~\cite{lemke}, between the Hamiltonian defined by Eq.(\ref{model})
and the Random Field Ising Model (RFIM).

We can rewrite the original Hamiltonian Eq.(\ref{model}) in the 
following way
\be
{\cal H}= - \sum_{<i_1 j_1>} \sigma_{i_1}  \sigma_{j_1} 
- \sum_{<i_2 j_2>} \tau_{i_2}  \tau_{j_2}  
- \sum_{i_1} \sum_{j_2(i_1)} J_{i_1 j_2} \sigma_{i_1} \tau_{j_2} \ ,
\label{ham_equiv}
\ee
where the indices $i_1, j_1$ ($i_2, j_2$) run over the sub-lattice 1
(respectively 2), $<i_1 j_1>$ ($<i_2 j_2>$) 
denotes sum to first nearest neighbors
pairs in the sub-lattice 1 (respectively 2)
and $j_2(i_1)$ denotes the sum over the two nearest neighbors ($j_2$'s) of
the site $i_1$ (in the two positive directions from $i_1$) 
in the whole lattice. 
Moreover we have denoted the variables of the
sub-lattice 1 (2) as $\sigma$'s (respectively $\tau$'s).

Now we fix the temperature to a small value  and all the spins inside the
sub-lattice 1 are fixed in the up state.  
Next we will examine the cost in energy to
do a compact droplet (in the sub-lattice 1) with all its spins flipped
down,  with the spins of the sub-lattice 2 fixed to an
arbitrary configuration.

The Hamiltonian of the model with all the $\tau$ spins
fixed to an arbitrary configuration is, modulo a constant,
\begin{eqnarray}
\nonumber
{\cal H}[\sigma | \tau \,\,{\rm fixed} ]&=& 
- \sum_{<i_1 j_1>} \sigma_{i_1}  \sigma_{j_1}   
- \sum_{i_1} \left[\sum_{j_2(i_1)} J_{i_1 j_2} \tau_{j_2} \right] 
\sigma_{i_1} \\
   &\equiv& - \sum_{<i_1 j_1>} \sigma_{i_1}  \sigma_{j_1}   
- \sum_{i_1} h_{i_1} \sigma_{i_1} \ ,
\end{eqnarray}
i.e. a Random Field Ising Model (RFIM) where the magnetic field 
 has zero mean and variance:
\be
\overline{h_i h_j}= \left\{ 
                \begin{tabular}{l l}
                 $2 \sigma^2_J$& if $i=j$ , \\
                 $0$            & elsewhere.
                \end{tabular}                 
                \right.
\ee

In particular we can think that all the spins in the $\tau$
sub-lattice are fixed up.  It is easy to reproduce all the steps of
the Imry-Ma argument~\cite{ma,NATTERMANN} (by turning on the temperature)
to show that one can find large regions where it is favorable
energetically to flip all the spins inside the region and hence to
destroy the long range order of the sub-lattice 1 independently of the
configuration of the sub-lattice 2 (and vice-versa).  

Thus we have
shown that the ferromagnetic order (in either or in both sub-lattices)
is unstable against an infinitesimal strength of the spin glass couplings:
i.e. initially the sub-lattices 1 and 2 are fixed to up (staggered
ferromagnetic order) and we have found that for any configuration
$\tau$ the sub-lattice 1 disorders, and by redoing the same steps with
sub-lattice 1 fixed (and disordered) we can see that the originally ordered
sub-lattice 2 disorders too.

Moreover, following Binder~\cite{BINDER}, we should expect that there
exists a crossover length $R_c$ such that for $R<R_c$ the staggered
ferromagnetic  order is stable but this order is unstable for  scales
$R>R_c$. The analytical expression for a RFIM with ferromagnetic
coupling $K$ and uncorrelated magnetic field with variance
$\sigma_h^2$ is~\cite{BINDER}
\begin{equation}
R_c \propto \exp \left[ C\left(\frac{K}{\sigma_h}\right)^2\right]\ ,
\end{equation}
where $C$ is a constant (O(1)), and so for our particular model 
(where $K=1$ and $\sigma_h^2=2 \sigma_J^2$), we finally obtain
\begin{equation}
R_c \simeq \exp \left[ \frac{C}{2 \sigma^2_J}\right]=
\exp \left[ \frac{C}{2 \lambda^2}\right] \ .
\end{equation}

Obviously when $\lambda$ is infinitesimally small (i.e. we put an
infinitesimal amount of spin glass disorder in the model) the
crossover ratio is exponentially large and ferromagnetic order will
only be destabilized in extremely large systems. Nevertheless, in the
thermodynamic limit the spin glass disorder is always relevant.

Finally, we can estimate that for $\lambda=0.5$ (the value that we
have used in our numerical simulations presented in this work) $R_c
\simeq 7$, assuming that $C$ is just 1.

At this point for $L > R_c$, where $L$ is the linear size of the
system, the picture is the following: both sub-lattices
are broken in clusters (inside of them all the spins, in average, point
in the same direction) of size less than $R_c$ interacting between
them. From the Imry-Ma argument it is clear that the
``effective'' interaction between these clusters is short range
(i.e. it could be very large but not infinite). Consequently, we have a two
dimensional spin glass with short range interactions. This phase  can be
thought as a frozen disordered phase (like the spin glass phase) where
the clusters play the role of the spin in the usual spin glass
phase. 

But we know that there exists no spin glass order at finite
temperature for short range spin glasses in two dimensions, and so we
conclude that there must be a second crossover from the spin glass
behavior to paramagnetic behavior as the size of the system
increases. We need only a correlation length greater than the range of
the interaction between the clusters to return to the usual short
range spin glass in two dimensions that has no phase transition.  In
the next sections we will try to put this fact in more quantitative
grounds.

%+++++++++++++++++++++++++++++++++++++++++++++++++++++++++++++++++++++++++++
\section{\label{N_RES}Numerical Simulations and Observables}

We have simulated, using the Metropolis algorithm, systems with linear sizes
ranging from $L=4$ to $L=48$ and averaging over 200 to 10000 samples
depending on the size. The largest lattice size simulated in reference 
\cite{lemke} used in the computation of their Binder cumulant was $L=12$.

In all the runs we have used an annealing procedure from higher
temperatures to the lower ones in order to thermalize the system. In
Table \ref{table:stat} we report the statistics we have used. We have
performed in the annealing procedure for all the temperatures the same
number of thermalization steps ($N_T$), that we have written in Table
\ref{table:stat}. 

For a given temperature we run $N_T$ steps for
thermalization and we
measure during $2 N_T$ and then we lower the temperature and we repeat
the process always with the same $N_T$. 
We will return to the issue of the thermalization time at
the end of this section.

\begin{table}[htbp]
\begin{center}
\begin{tabular}{|c|c|c|c|} \hline
$L$       & $T$'s & samples & Termalization time   \\ \hline
4       & [1.5,4.0] & 10000 &10000     \\ \hline
6       & [1.5,4.0] & 10000 &10000     \\ \hline
8       & [1.5,4.0] & 4000 &30000   \\ \hline
16      & [1.5,3.0] & 1633  &30000   \\   \hline
24      & [1.5,3.0] & 700   &60000   \\   \hline
32      & [1.5,3.0] & 1576  &90000   \\   \hline
48      & [1.4,2.5] & 426(*)   &900000 \\ \hline
\end{tabular}
\end{center}
\caption{ Description of our runs. The step in temperatures was $0.1$
in all  runs. (*) means that for the $L=48$ lattice in the measure of
the staggered overlap we only have simulated 100 samples.}
\label{table:stat}
\end{table}

In the next paragraphs we will describe the observables that we have
measured in our numerical simulations.

We have measured the global ($m$) and staggered ($m_s$) magnetizations :
\be
m\equiv \frac{1}{L^2} \sum_i S_i  \,\,\, , \,\,\,  m_s\equiv \frac{2}{L^2}
\sum_{i_1}  S_{i_1} \ ,
\label{magnetization}
\ee
where the sum runs over all the lattice sites ($i$) and over a sub-lattice 
($i_1$) respectively. 

In order to calculate spin glass quantities we have simulated two
replicas $\alpha$ and $\beta$ in parallel with the same disorder. 
The overlaps between the replicas, global ($q$) and staggered ($q_s$),
are:
\be
q \equiv \frac{1}{L^2} \sum_i S_i^{\alpha} S_i^{\beta} \,\,\, , \,\,\,
q_s \equiv \frac{2}{L^2} \sum_{i_1} S_{i_1}^{\alpha} S_{i_1}^{\beta} \ ,
\label{overlap}
\ee
where, again, the sum runs over all the lattice sites ($i$) and over one of
the two sub-lattices ($i_1$) respectively.

The magnetic global and staggered susceptibilities (without the
$\beta$ factor) are defined as:
\be
\chi \equiv L^2 \left[\overl{\lan m^2 \ran}- \left(\overl{\lan |m|
\ran}\right)^2\right] 
\,\,\, , \,\,\,
\chi_s \equiv \frac{L^2}{2} \left[\overl{\lan m_s^2 \ran} - \left(\overl{\lan |m_s|
\ran}\right)^2\right]\ .
\label{sus_m}
\ee

The spin glass or overlap susceptibilities ($\chi_q$, $\chi^s_q$) 
are defined by:
\be
\chi_q \equiv L^2 \left[\overl{\lan q^2 \ran} - \left(\overl{\lan |q|
\ran}\right)^2\right] \,\,\, , \,\,\,
\chi_q^s \equiv \frac{L^2}{2}\left[ \overl{\lan q_s^2 \ran}
- \left(\overl{\lan |q_s|
\ran}\right)^2\right] \ .
\label{sus_q}
\ee

We have measured also the Binder cumulants of the 
magnetization: global $g_m$ and staggered $g_m^s$, 
\be
g_m \equiv \frac{1}{2} \left[ 3-\frac{\overl{\lan m^4 \ran}}{\left(\overl{\lan
m^2 \ran}\right)^2} \right] 
\,\,\, , \,\,\,
g_m^s \equiv \frac{1}{2} \left[ 3-\frac{\overl{\lan m_s^4 \ran}}{\left(\overl{\lan
m_s^2 \ran}\right)^2} \right] \  
\label{binder_m}
\ee
and Binder parameters of the overlaps: global $g_q$ and staggered $g^s_q$,
\be
g_q \equiv \frac{1}{2} \left[ 3-\frac{\overl{\lan q^4
\ran}}{\left(\overl{\lan q^2 \ran}\right)^2} \right]  
\,\,\, , \,\,\,
g_q^s \equiv \frac{1}{2} \left[ 3-\frac{\overl{\lan q_s^4
\ran}}{\left(\overl{\lan  q_s^2 \ran}\right)^2} \right]  \ .
\label{binder_q}
\ee

Finally the specific heat is defined by:
\be
C_V \equiv \frac{1}{L^2} \left(\overl{\lan {\cal H}^2 \ran} - \overline{\lan
{\cal H}
\ran}^2 \right) \ .
 \label{cesp}
\ee

\begin{figure}
\begin{center}
\addvspace{1 cm}
\leavevmode
\epsfysize=250pt
\epsffile{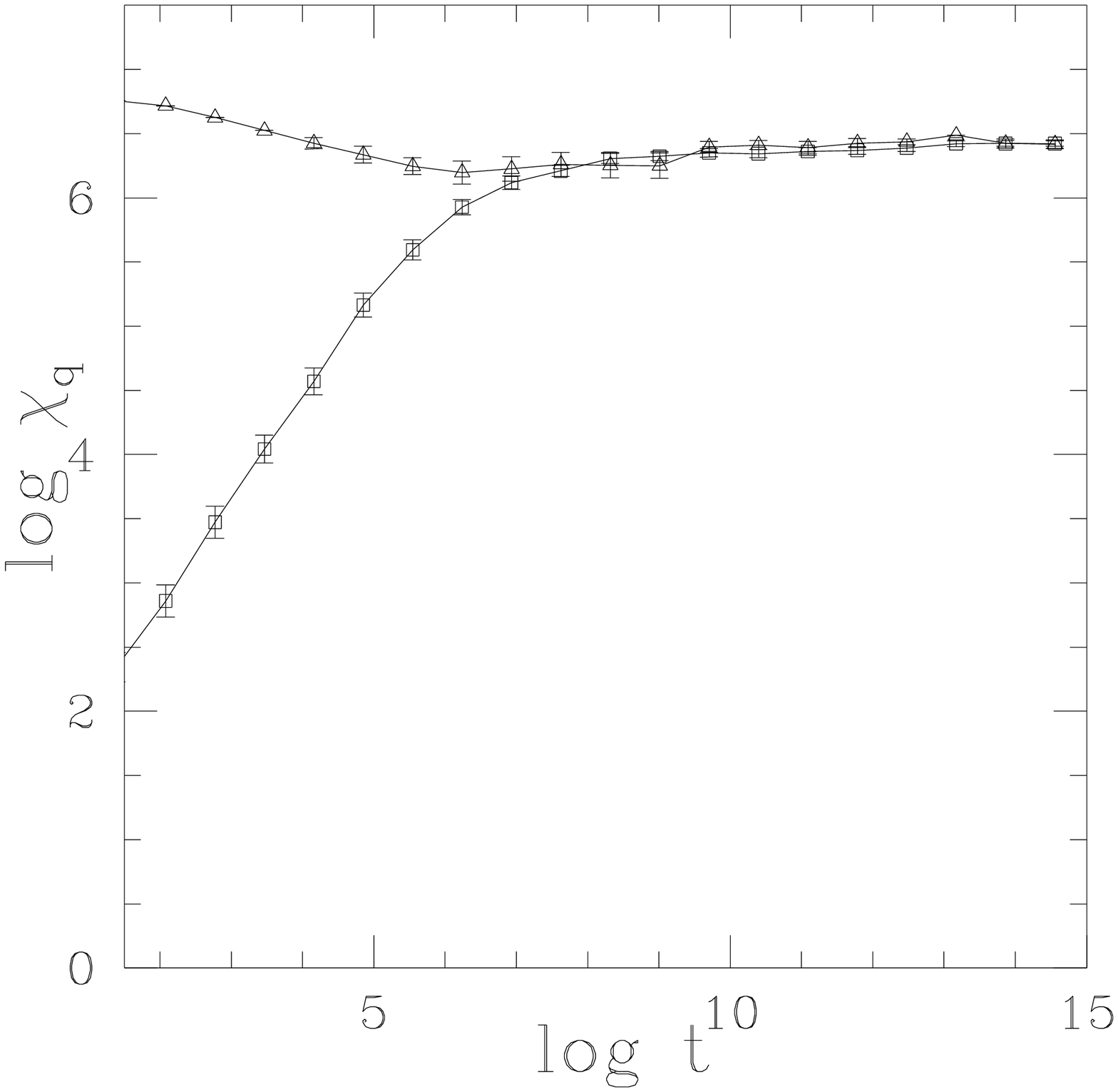}
\end{center}
\caption{The (non connected) 
overlap susceptibility against the Monte Carlo time in a
double logarithmic scale for one of the lower temperature that we have
simulated ($T=1.5$) and for $L=32$. The number of samples was 100. 
The upper curve is from a configuration with all the spins up and the
lower curve is with the starting configuration chosen random. One can
say that the system has thermalized when there is no difference between
the two curves and both stay on a plateau. We have chosen as
thermalization time $t=90000$ (i.e. $\log t=11.4$) .}
\label{fig:terma}
\end{figure}

\begin{figure}
\begin{center}
\addvspace{1 cm}
\leavevmode
\epsfysize=250pt
\epsffile{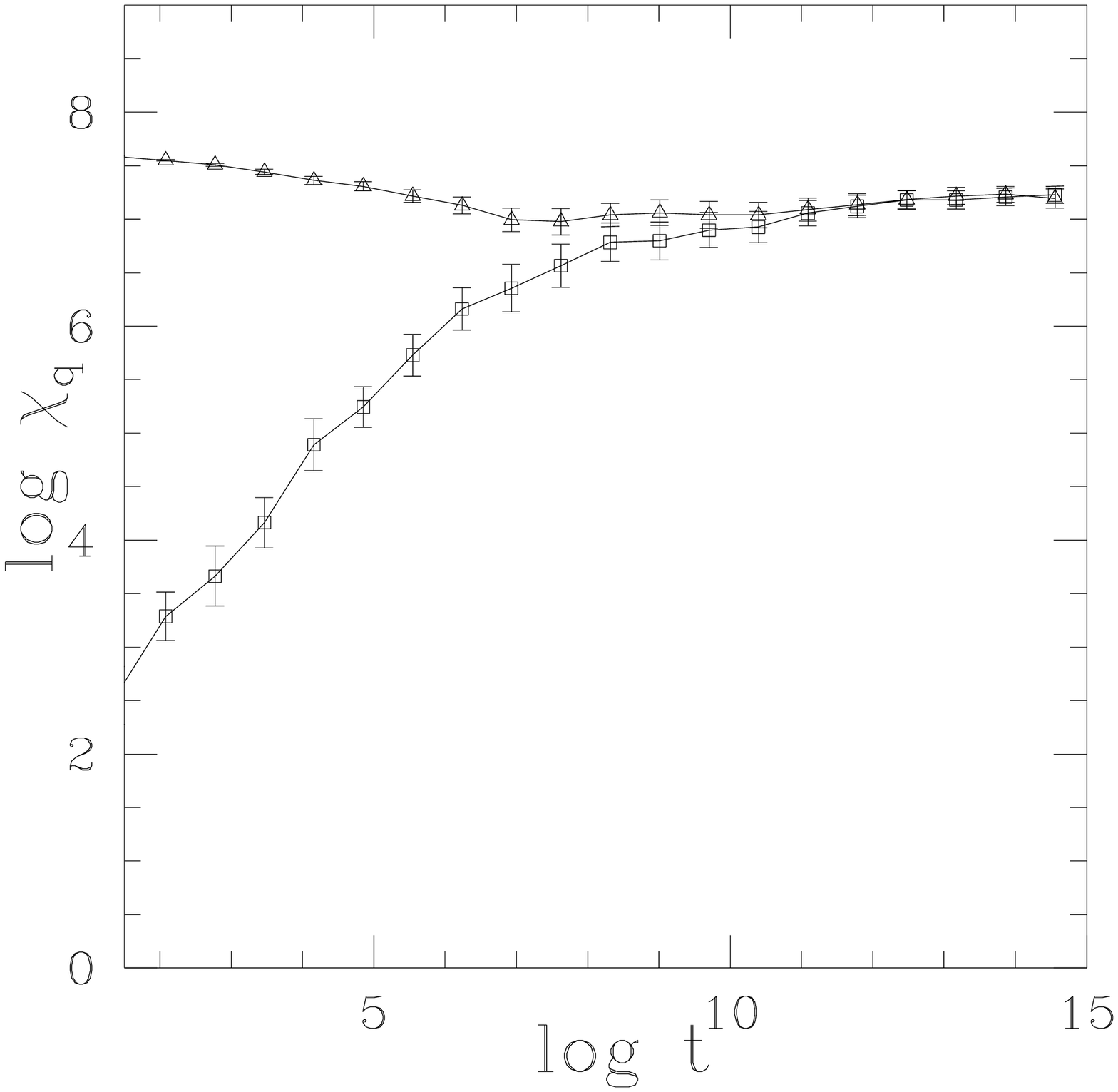}
\end{center}
\caption{The (non connected) 
overlap susceptibility against the Monte Carlo time in a
double logarithmic scale for  the lowest temperature that we have
simulated ($T=1.4$) and for the largest lattice $L=48$. 
The number of samples was 25. 
The upper curve is from a configuration with all the spins up and the
lower curve is with the starting configuration chosen random.  
We have chosen as
thermalization time $t=900000$ (i.e. $\log t=13.7$) .}
\label{fig:terma2}
\end{figure}

In order to decide a safe thermalization time, $N_T$, (written in
Table \ref{table:stat}) we have used the method proposed by Bhatt and
Young~\cite{bhatt} which consists in running, at the lowest
temperature, with an ordered (all spins up) and high temperature
initial configurations and monitoring the behavior of the
susceptibilities (in our case the non connected 
overlap susceptibility: $L^2 \overline{\lan q^2 \ran}$) with the
Monte Carlo time. When the two curves reach the same plateau we can
say that the system has thermalized.  We show in Figures
\ref{fig:terma} and \ref{fig:terma2} this procedure for two of our
biggest lattices and at the lower temperatures simulated (i.e. $L=32$
and $T=1.5$ and $L=48$ and $T=1.4$ respectively).

We remark that we have used for {\em all} temperatures of the
annealing procedure the value that we have computed for the lower one
(that we have reported in Table \ref{table:stat}). 

\section{Numerical Results}

\subsection{Crossover Ferromagnetic-Spin Glass}

In this sub-section we will show numerical evidences of the first
crossover: from a staggered ferromagnetic phase to a ``spin glass'' phase. 

\begin{figure}
\begin{center}
\addvspace{1 cm}
\leavevmode
\epsfysize=250pt
\epsffile{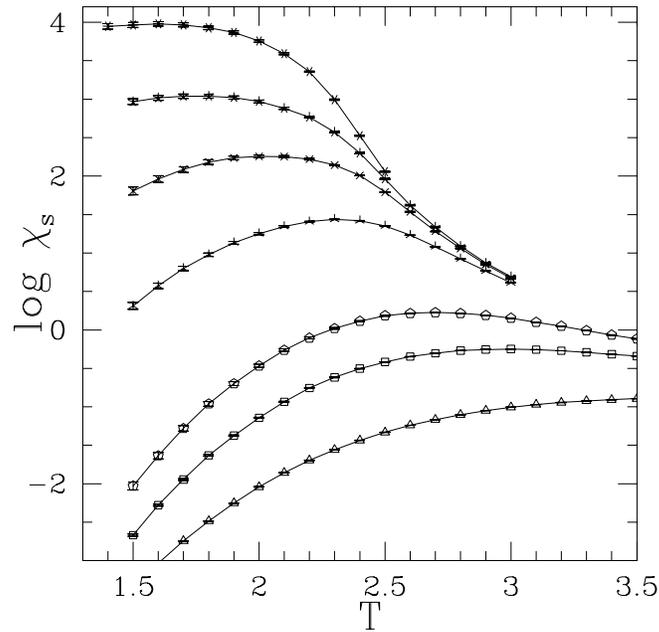}
\end{center}
\caption{Susceptibility of the staggered magnetization as a 
function of the temperature. The lattice sizes are (bottom to top): 
4, 6, 8, 16, 24, 32 and 48. Here and in the rest of the figures we have used
the following symbols for the lattices: 
$L=4$ triangle, $L=6$ square, $L=8$ pentagon,
$L=16$ three-line star, $L=24$ four-line star, $L=32$ five-line star and 
$L=48$ six-line star.}
\label{fig:sus_ms}
\end{figure}

\begin{figure}
\begin{center}
\addvspace{1 cm}
\leavevmode
\epsfysize=250pt
\epsffile{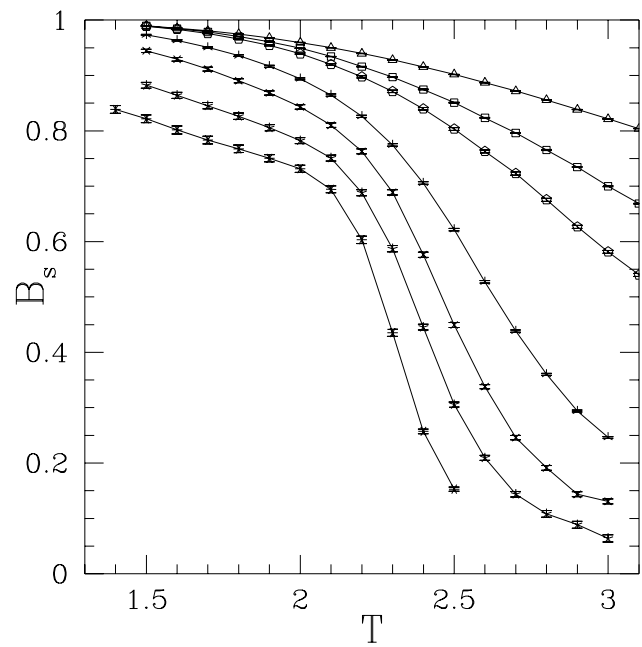}
\end{center}
\caption{Binder cumulant of the staggered magnetization as a 
function of the temperature. The lattice sizes are (bottom to top in
the right part of the plot): 48, 32, 24, 16, 8, 6 and 4.}
\label{fig:binder_ms}
\end{figure}

It is natural to study this crossover examining firstly the
susceptibility and the Binder cumulant of the staggered magnetization,
Figures \ref{fig:sus_ms} and \ref{fig:binder_ms} respectively.  They
are the observables that at $\lambda=0$ describe the
paramagnetic-staggered ferromagnetic phase transition.

In Figure \ref{fig:sus_ms} it is possible to see how the point where the
staggered susceptibility reaches the maximum drifts quickly to
zero with increasing system size. 
Moreover in Figure \ref{fig:binder_ms} there is no crossing of
the Binder cumulant: this is a stronger evidence that this parameter
(the staggered magnetization) does not show a phase transition at
finite temperature, i.e. in the thermodynamic limit $\lan |m_s|
\ran=0$ for all temperatures different from zero. 
 
\begin{table}
\begin{center}
\begin{tabular}{|c|c|c|} \hline
L       & $T(\chi_s^{\rm max})$ &$\chi_{s}^{\rm max}$   \\ \hline
8       & 2.7(1)        & 1.255(4)  \\ \hline
16      & 2.3(1)        & 4.21(2)  \\   \hline
24      & 2.0(1)        & 9.6(2)   \\    \hline
32      & 1.8(2)        & 19.9(4)  \\   \hline
32      & 1.6(2)        & 54(2)  \\   \hline
\end{tabular}
\end{center}
\caption{Maximum of $\chi_s(L,T)$ in temperature
and the temperature at which $\chi_s(L,T)$
reaches the maximum.}
\label{table:staggered}
\end{table}

More quantitatively, from the data of Table \ref{table:staggered} it is
clear that the temperatures in which $\chi_s$ reaches the
maximum, that we denote as $T(\chi_s^{\rm max}) $, goes to zero following a
power law: 
\be
T(\chi_s^{\rm max}) \propto L^{-0.29(4)} ,
\ee 
using only the data
of the lattices: 8, 16, 24, 32 and 48. This fit has $\chi^2/{\rm DF}=0.5/3$,
where DF means for the number of degrees of freedom in the fit.

This numerical results support our previous analytic result
that the staggered ferromagnetic phase is unstable
against an infinitesimal perturbation of the kind of a spin glass
interaction between the two sub-lattices.

\begin{figure}
\begin{center}
\addvspace{1 cm}
\leavevmode
\epsfysize=250pt
\epsffile{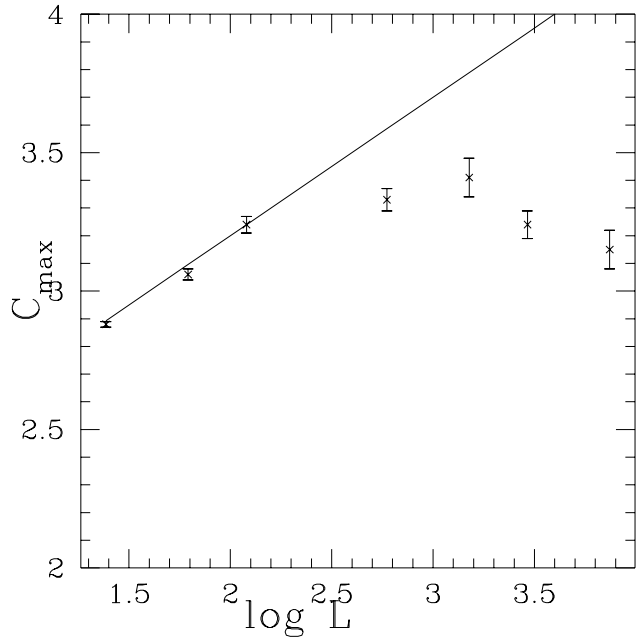}
\end{center}
\caption{The maximum specific heat as a function of $L$ for
$L=4,6,8,16,24, 32$ and $48$. 
We have marked the finite size scaling  prediction
of a logarithmic divergence for the pure model ($\lambda=0$). 
The straight line has been obtained from a fit
to the formula $A \log L +B$ using $L=4,6$ and $8$.}
\label{fig:c_l}
\end{figure}

Moreover, we have predicted the presence of a crossover between the
staggered ferromagnetic 
phase and a ``spin-glass'' like phase for lattices of linear sizes of order
$10$. To see this crossover we can analyze the specific heat. 
All the lattice sizes show a maximum near the Onsager
temperature ($T_c=2.26$). 
We will see that these maxima are the ``souvenir'' of the staggered
phase transition. In particular we can examine the scaling of
the maxima, that we denote $C_{\rm max}(L)$, against the lattice size. The
result is given in Figure \ref{fig:c_l}. In this figure we have also
plotted the finite size scaling prediction for the pure Ising model
($\lambda=0$) that is
\be
C_{\rm max}(L) \propto \log L \ .
\ee

It is clear from Figure \ref{fig:c_l} that up to $L=8$ the data follow
in good agreement the prediction of the pure model (i.e. up to $L=8$
the low temperature region is staggered ferromagnetic). But between
$L=8$ and $L=16$ the system crosses over to a different behavior where
there is no divergence of the specific heat, i.e. $\alpha<0$.  This
result is in very good agreement with our analytical estimate $R_c
\simeq 7$.

Hence, this plot shows us clearly a first crossover between the
staggered ferromagnetic phase and a ``spin glass'' like phase. In the first
part of the crossover the specific heat diverges logarithmically
($\alpha=0$) whereas in the second part of the crossover the specific
heat does not diverge ($\alpha<0$).

It is clear that the task that remains is to check that effectively
what we have named as a ``spin glass'' phase has really the properties
of a spin glass phase.

To do this we can study the susceptibility and the Binder parameter of
the total overlap. In a spin glass phase the magnetization is zero but
not the overlap, that becomes the order parameter. If the low
temperature phase is spin glass the overlap susceptibility should peak
near the transition temperature and the value of the peak should grow with
some power of the lattice size (more precisely as
$L^{\gamma/\nu}$). Moreover, the analysis of the Binder cumulant should
show a clear crossing between curves of different lattice sizes.

\begin{figure}
\begin{center}
\addvspace{1 cm}
\leavevmode
\epsfysize=250pt
\epsffile{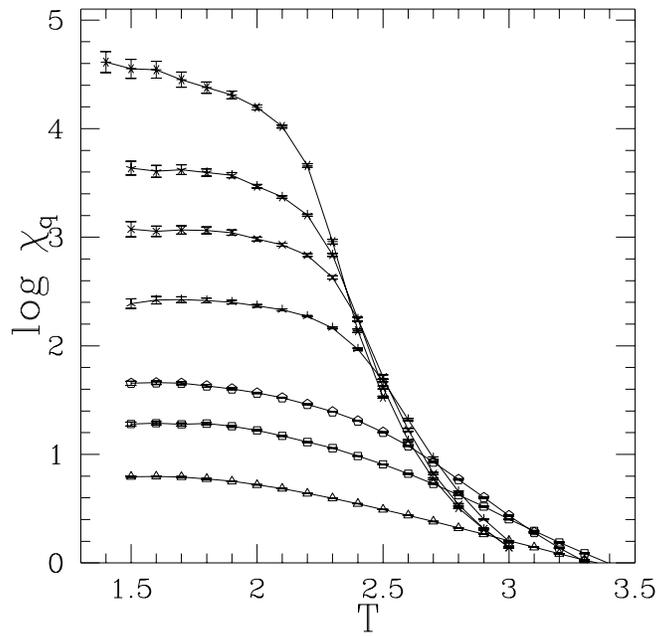}
\end{center}
\caption{Susceptibility of the total overlap against the temperature. 
The lattices sizes are (top to bottom): 48, 32, 24, 16, 8, 6 and 4. 
From this figure it is clear that $\chi_q$ for $L=16, 24, 32$ and $48$
does not show a maximum.}
\label{fig:sus_q}
\end{figure}

\begin{figure}
\begin{center}
\addvspace{1 cm}
\leavevmode
\epsfysize=250pt
\epsffile{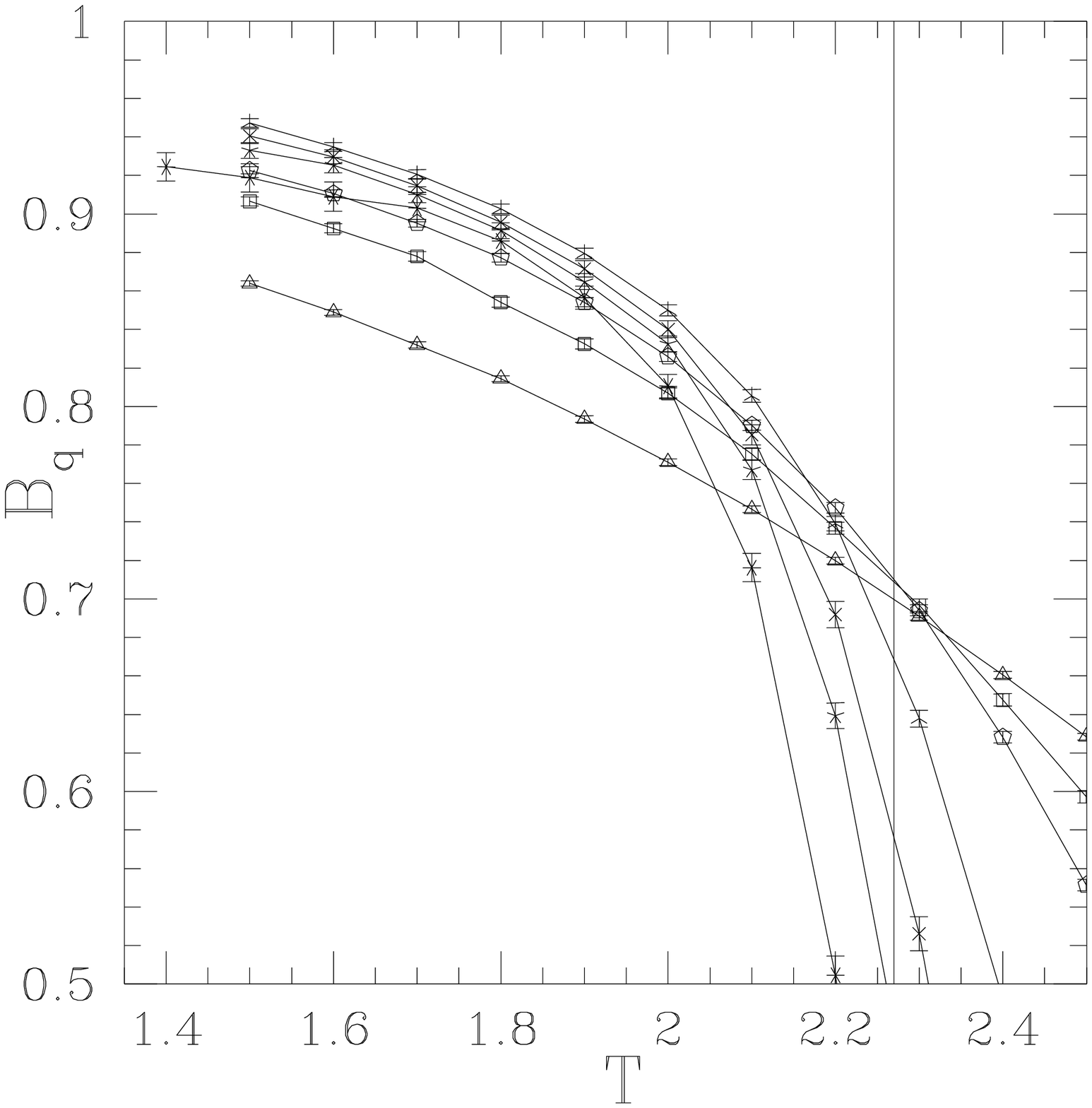}
\end{center}
\caption{Binder parameter of the total overlap against the temperature. 
The lattices sizes are (bottom to top in the right part of the plot): 
48, 32, 24, 16, 8, 6 and 4. We have marked with a vertical line the
Onsager temperature.}
\label{fig:binder_q}
\end{figure}

In Figures \ref{fig:sus_q} and \ref{fig:binder_q} we show the data for
the susceptibility and Binder parameter of the total overlap.
We can see again the crossover staggered-spin glass in Figure
\ref{fig:binder_q}. The curves of the lattice sizes $4$, $6$ and $8$
cross practically at the critical temperature of the pure model
($\lambda=0$, vertical line). If we examine the crossing point of
pairs of lattices for growing sizes, we observe that these crossing
points drift to lower temperatures. For instance, the crossing point
of the $16$ and $32$ lattices is near $T=1.9$.  This suggests that the
limit of this sequence of crossing points may be zero. This would
imply that the spin glass phase is unstable and is crossing over to a
paramagnetic phase.

Moreover, the overlap susceptibility does not present a sharp  maximum
as a function of the temperature (see the $L=24, 32$ and $48$ curves). 
We can extract one conclusions of this fact:

\begin{itemize}

\item As long as the spin glass is stable we will expect again that
$\chi_q(T)$ should show a sharp maximum as a function of $T$, or at
least that we have independent spin glass order in both sub-lattices
(i.e. some sort of staggered spin glass order).  From figure
\ref{fig:sus_q} it is clear that $\chi_q$ does not show a sharp peak
in the region that we have simulated (see, for instance, the $L=32$
and $L=48$ data).  Moreover the crossing point of the Binder cumulant of the
total overlap of two different lattices is drifting to lower values of
the temperature, and so we pass to discus the second possible option:
spin glass order on both sub-lattices.  In this case the staggered
overlap susceptibility should have a sharp peak at an intermediate
temperature which characterizes a phase transition between a
paramagnetic phase and a spin glass one.

\end{itemize}

To study this issue in more detail, we will examine in the next
subsection the overlap defined only in one of the two sub-lattices.

\subsection{Crossover Spin Glass-Paramagnetic}

\begin{figure}
\begin{center}
\addvspace{1 cm}
\leavevmode
\epsfysize=250pt
\epsffile{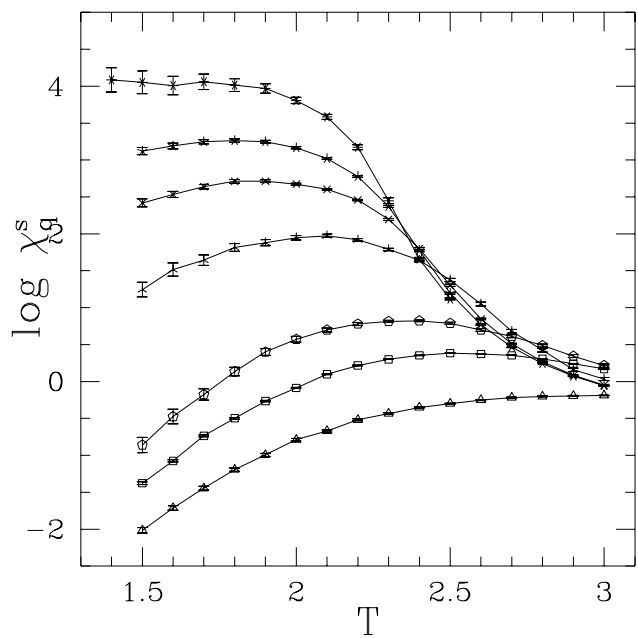}
\end{center}
\caption{Susceptibility  of the staggered  overlap against the temperature. 
The lattices sizes are (bottom to top): 48, 32, 24, 16, 8 and 4.}
\label{fig:sus_qs}
\end{figure}

\begin{figure}
\begin{center}
\addvspace{1 cm}
\leavevmode
\epsfysize=250pt
\epsffile{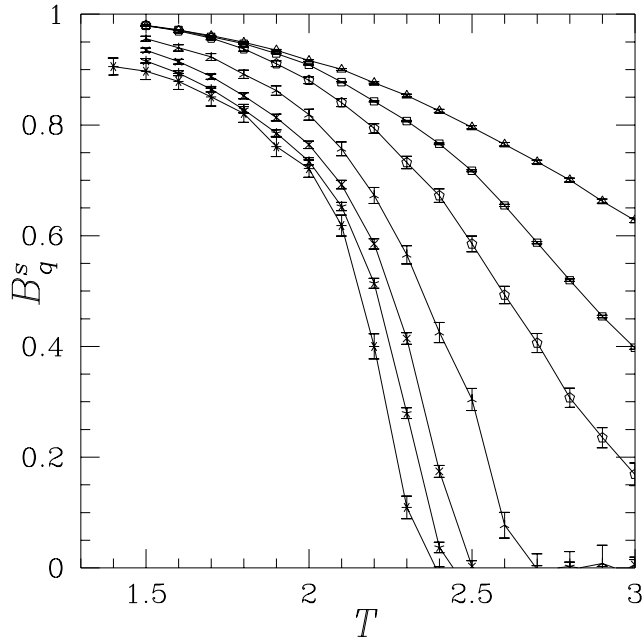}
\end{center}
\caption{Binder parameter of the staggered  overlap against the temperature. 
The lattice sizes are (bottom to top): 48, 24, 32, 16, 8, and 4.}
\label{fig:binder_qs}
\end{figure}

In Figures \ref{fig:sus_qs} and \ref{fig:binder_qs}
we show the susceptibility and the Binder cumulant of the staggered
overlap (i.e. the overlap computed only in one of the two sub-lattices).

It is clear that these two figures are similar to Figures \ref{fig:sus_ms} and 
\ref{fig:binder_ms}. For example, there is no crossing in the Binder
cumulant. We remark that we have performed small statistic on the
$L=48$ lattice and so we obtain large errors. Taking into account only the
lattice sizes $L=4,6,8,16,24$ and 32 the effect is clear: the
thermodynamical Binder cumulant goes to zero for all the temperatures
simulated.

\begin{table}
\begin{center}
\begin{tabular}{|c|c|c|} \hline
L         &  $T(\chi_{qs}^{\rm max})$ & $\chi_{qs}^{\rm max}$     \\ \hline
8       & 2.4(1)        & 2.27(3) \\ \hline
16      & 2.1(1)        & 7.2(2) \\   \hline
24      & 1.9(1)        & 15.1(3) \\ \hline
32      & 1.6(2)        & 28(2)   \\   \hline
\end{tabular}
\end{center}
\caption{Maximum in temperature of the staggered overlap 
susceptibility $\chi_q^s(L,T)$, 
and the temperature at which $\chi_q^s(L,T)$ reaches the maximum. For
$L=48$ the error bars do not permit a safe estimate of the maximum.}
\label{table:q-staggered}
\end{table}
The ``apparent'' critical temperature (defined as the value where
$\chi_q^s$ reaches the maximum) goes to zero following the law
\be
T(\chi_{qs}^{\rm max}) \propto L^{-0.23(5)}
\label{scaling}
\ee
where we have used L = 8, 16, 24 and 32 in the fit that has
$\chi^2/{\rm DF}=0.8/2$. The data used in this fit has been reported
in Table \ref{table:q-staggered}.

Within the statistical error the exponent is just the same as in the
staggered magnetization. We can compare this figure with that computed
for the 2D spin glass\cite{bhatt}: $1/\nu=0.38(6)$. If we are seeing
the pure two dimensional spin glass transition which occurs at $T=0$,
then we will expect a behavior of the apparent critical temperature
like $T_{\rm app} \propto L^{-0.38(6)}$, which is, within the
statistical error, the law that we have found for the present model
(Eq. (\ref{scaling}))\footnote{ The shift of the apparent critical
temperature follows a law: $T_c(L)-T_c \propto L^{-1/\nu}$.}: the
difference between the two exponents is $0.15(8)$ (i.e. almost two
standard deviations). Obviously our simulations were done in a range of
temperatures far away of the critical point ($T=0$).

Both Figures \ref{fig:sus_qs} and \ref{fig:binder_qs} suggest that the
phase transition is at $T=0$. In other words, there is not a spin
glass phase in the sub-lattices 1 and 2. The whole system seems to be
crossing over to a paramagnetic phase.

%+++++++++++++++++++++++++++++++++++++++++++++++++++++++++++++++++++++++++++
\section{\label{S_CONCLU}Conclusions}

We have studied the two dimensional Ising spin glass model with next
nearest neighbor interactions both
analytical and  numerically.

We have analytically obtained that the system should present two
different crossovers as the volume grows: the first one from a
staggered ferromagnetic phase to a spin glass phase for intermediate
sizes, and a second crossover between this spin glass phase to a
paramagnetic one that will dominate the physics in the thermodynamic
limit. Moreover we have obtained the dependence of the first crossover
length (staggered-spin glass) on the parameters of the system.

Then we have checked this scenario by performing extensive numerical
simulations. We have clearly established the first crossover
(staggered-spin glass) and have found strong evidences of the second (spin
glass-paramagnetic). In particular, the exponent that governs the
shift of the maxima of the spin glass susceptibility is compatible
with the known value for the pure two dimensional spin glass.

\vspace{2cm}
{\large \bf {\label{S_ACKNOWLEDGES}Acknowledgments}}

We have run mainly using ALPHA Workstations but the large lattices
were run in the RTNN parallel machine\footnote{A parallel machine
composed by 32 Pentium-Pro  at Zaragoza University.}. 
We acknowledge the RTNN project for
permitting to us the use of the RTNN computer.

J.~J.~Ruiz-Lorenzo is supported by an EC HMC (ERBFMBICT950429)
grant and thanks to L.A. Fern\'andez for interesting discussions. 
D. A. Stariolo is partially supported by Conselho Nacional de
Desenvolvimento Cientifico e Tecnol\'ogico (CNPq), Brazil.


\newpage

\begin{thebibliography}{99}

\bibitem{BOOK} E. Marinari, G. Parisi and J. J. Ruiz-Lorenzo,
``Numerical Simulations of Spin Glass Systems'', in {\it ``Spin
Glasses and Random Fields"}, edited by A. P. Young (World Scientific,
Singapore, 1997), 130.

\bibitem{dekker}C. Dekker, A.F.M. Arts, H.W. de Wijn, A.J. van
Duyneveldt and J.A. Mydosh, {\it Phys.Rev.Lett.} {\bf 61}, 1780
(1988);\\ C. Dekker, A.F.M. Arts and H.W. de Wijn, {\it Phys.Rev.}
{\bf B38}, 8985 (1988);\\ C. Dekker, A.F.M. Arts, H.W. de Wijn,
A.J. van Duyneveldt and J.A. Mydosh, {\it Phys.Rev.} {\bf B40}, 11243
(1989).

\bibitem{mydosh} J. A. Mydosh, {\it Spin Glasses: an Experimental
 Introduction}, (Taylor \& Francis, London 1993).

\bibitem{review} H. Rieger, {\em Annual Reviews of Computational
Physics II} (World Scientific, Singapore 1995), p. 295.

\bibitem{bhatt} R.N. Bhatt and A.P. Young, {\it Phys. Rev.} {\bf B37},
5606 (1988).

\bibitem{kawashima} N. Kawashima, N. Hatano and M. Suzuki, {\it J.
Phys.} {\bf A25}, 4985 (1992).

\bibitem{rieger} H. Rieger, B. Steckemetz and M. Schreckenberg, {\it
Europhys. Lett.} {\bf 27}, 485 (1994);\\ H. Rieger, L. Santen,
U. Blasum, M. Diehl and M. J\"unger, {\it J.  Phys.} {\bf A29}, 3939
(1996);\\ N. Kawashima and H. Rieger, {\it Europhys. Lett.} {\bf 39},
85 (1997).

\bibitem{SHIRAKURA} T. Shirakura and F. Matsubara, {\it Phys. Rev. Lett.}
{\bf 79}, 2755 (1997).


\bibitem{lemke} N. Lemke and I. A. Campbell, {\it Phys. Rev. Lett.}
{\bf 76}, 4616 (1996).

\bibitem{ma}Y. Imry and S. K. Ma, {\it Phys. Rev. Lett.} {\bf 35},
1399 (1975).

\bibitem{NATTERMANN} T. Nattermann, ``Theory of the Random Field Ising
Model'', in {\it ``Spin Glasses and Random Fields"}, edited by
A. P. Young (World Scientific, Singapore 1997).

\bibitem{BINDER} K. Binder, {\it Z. Phys.} {\bf B50}, 343 (1983).

\end{thebibliography}
\end{document}